\documentclass[aps,groupedaddress,floatfix,superscriptaddress,amsmath,amssymb,twocolumn,pre, 10pt]{revtex4-2}

\usepackage{graphicx}
\usepackage{amsthm}
\usepackage{amsfonts}
\usepackage{bm}
\usepackage{amsmath}
\usepackage{upgreek}
\usepackage[version=4]{mhchem}

\usepackage{textcomp} 		
\usepackage{ifthen} 		
\usepackage{xifthen} 		
\usepackage{soul} 			
\usepackage{lineno}
\usepackage[normalem]{ulem}
\usepackage{siunitx}
\usepackage{xcolor} 
\usepackage{enumitem}
\AtBeginDocument{\RenewCommandCopy\qty\SI}
\usepackage{physics}
\usepackage[colorlinks=true,citecolor=blue,linkcolor=blue,urlcolor=blue, linktocpage=true,breaklinks=true,pagebackref=false]{hyperref}

\begin{document}

\title{Topography-based navigation in a millikelvin scanning tunneling microscope using binary-encoded position markers}

\author{Ronja Fischer-Süßlin}
\affiliation{Department of Physics, University of Konstanz, 78457 Konstanz, Germany}
\author{Roman Hartmann}
\affiliation{Department of Physics, University of Konstanz, 78457 Konstanz, Germany}
\author{Timo Kandra}
\affiliation{Department of Physics, University of Konstanz, 78457 Konstanz, Germany}
\author{Elke Scheer}
\affiliation{Department of Physics, University of Konstanz, 78457 Konstanz, Germany}

\begin{abstract}
We present a compact millikelvin scanning tunneling microscope (STM) operating at 270\,mK 
with topographic navigation to micron-scale targets. 
Two piezoelectric low-temperature nanopositioners extend the accessible sample area, while a multi-stage copper powder and capillary filter scheme preserves millikelvin energy resolution, verified by BCS spectroscopy on aluminium thin films. 
A lithographically fabricated binary-encoded gold pattern encodes unique 16-bit coordinates in 4x4 pixels of 200\,nm$\times$200\,nm each. We demonstrate absolute positioning across a 350x350\,\textmu m$^2$ area from a single STM scan. Requiring only a single lithography step and no hardware modifications to existing STM setups, the navigation system provides a versatile platform for scanning tunneling spectroscopy of nanoflakes and nanoscale devices.
\end{abstract}

\maketitle

\section{INTRODUCTION}
Scanning tunneling microscopy (STM) and spectroscopy (STS) at millikelvin temperatures provide access to low-energy excitations on the atomic scale. Examples of such excitations include quasiparticles in superconductors \cite{dibernardoSignatureMagneticdependentGapless2015, dieschCreationEqualspinTriplet2018, yazdaniProbingLocalEffects1997, ortuzarYuShibaRusinovStatesTwodimensional2022}. 
Their study requires to avoid magnetic components of the STM head and simultaneously the ability to be operated in variable applied magnetic fields.\\
Achieving millikelvin temperatures and high energy resolution with an STM is possible with a compact design that also operates in magnetic fields, as demonstrated in \cite{debuschewitzCompactVersatileScanning2007}. 
Due to the absence of optical access, limited scan range, and thermal drift, precisely localizing micron-scale two-dimensional devices in low-temperature STM remains challenging. This has largely confined STS investigations to thin films and extended layers, rather than discrete micron-scale structures or devices - motivating the need for navigation approaches that are compatible with low-temperature operation and high energy resolution. 
The absence of optical access can be partially compensated by combining STM with atomic force microscopy (AFM). Such a dual-mode instrument operating at 30\,mK \cite{lesueurPhaseControlledSuperconducting2008, sueurCryogenicAFMSTMMesoscopic2007} uses the AFM mode to locate and image nanocircuits before performing tunneling spectroscopy. The coarse positioning is done with inertia slip-stick motors in all three spatial directions. However, the added AFM functionality introduces considerable complexity to the system. 
\\
Another method to estimate the tip position over micron-scale conductive samples on insulating substrates is based on the effective tip-sample capacitance. By distinguishing insulating from conductive regions through capacitance changes on the order of $10^{-18}$\,F \cite{kolbCapacitiveSensorMicropositioning1998, liSelfnavigationScanningTunneling2011, molina-mendozaNoteLongrangeScanning2014}, crash-free navigation is possible while keeping the tip well above the surface. However, this method does not yield an absolute tip position on the conductive area. 
\\
Here, we present a topography-based navigation approach to locate micron-sized structures on a conductive substrate, extending the concept introduced in \cite{chenDesignFabricationGuiding2026} with a simpler and more universal sample design. Our design is inspired by \cite{coissardManybodyGroundStates2021} who used a markerfield to guide the tip to a graphene device in a combined AFM/STM \cite{coissardImagingTunableQuantum2022}. 
\\
We describe the integration of two piezoelectric nanopositioners into a millikelvin STM - including their cabling and the filtering concept required for low-temperature operation - which extends the accessible sample area while preserving energy resolution. We introduce a binary-encoded coordinate pattern which allows to navigate to arbitrary positions. This way, multiple devices or nanostructures can be placed and investigated on a single sample, which to our knowledge has not been demonstrated in other systems.

\section{EXPERIMENTAL SETUP}
\begin{figure}[htbp]
	\centering
	\includegraphics[width=0.7\linewidth]{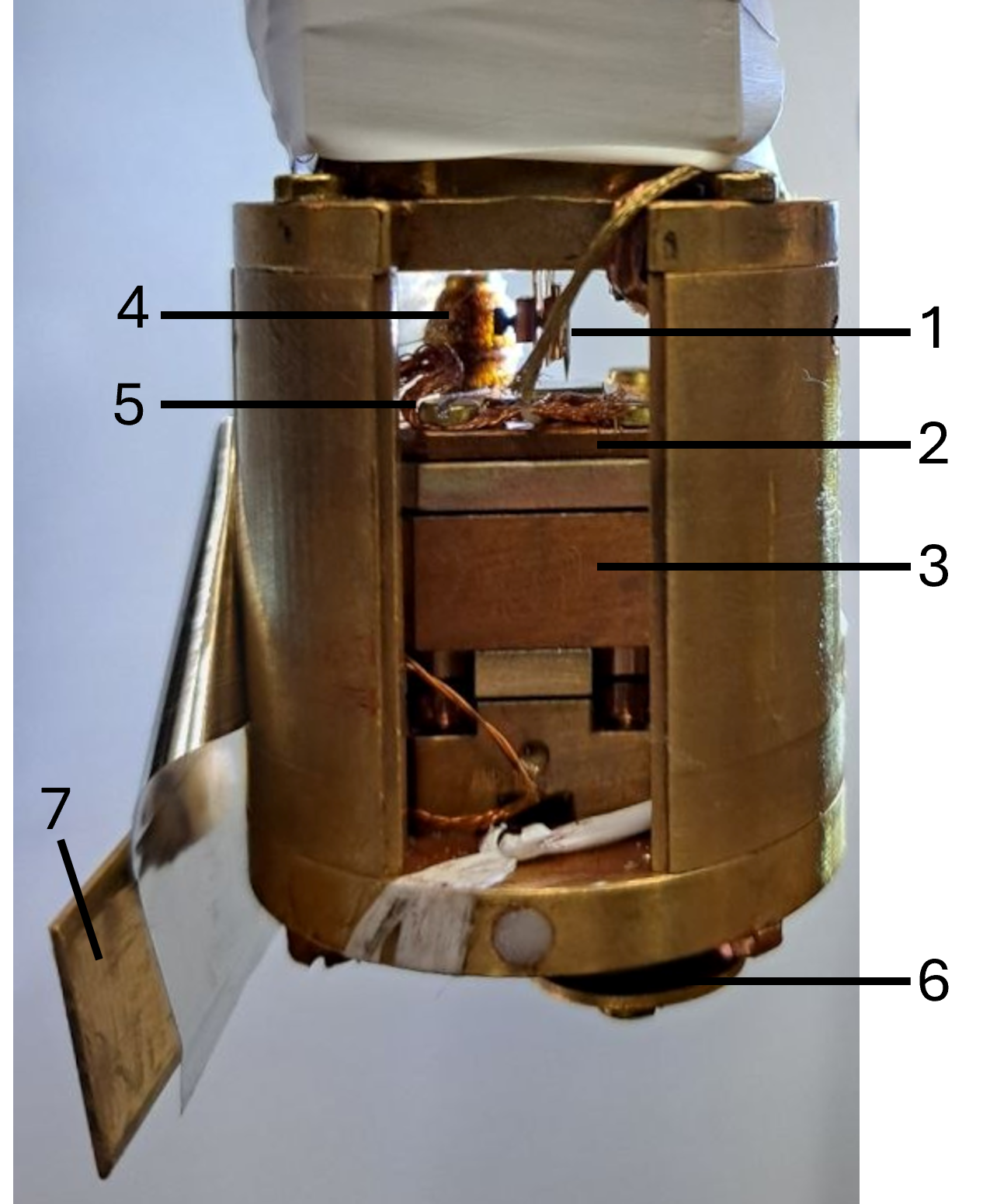}
	\caption{Photograph of the STM head showing (1) the copper slider with tip and (2) the sample plate on top of the (3) two linear nanopostioners. A sample heater (4) prevents adsorption on the sample, and for thermal coupling the sample plate is connected to the gold-plated copper shielding (5). Additionally to the thermometers of the cryostat insert, we mounted a sample thermometer (6). For radiation shielding, the STM head is closed with (7) small copper elements.}
	\label{fig:stm_setup}
\end{figure}

An important aspect in the design of the whole set-up is the limited space and the limited cooling power of the cryostat insert. 
We use a low temperature, nonmagnetic STM in a conventional liquid $^3$He cryostat (Heliox VL by Oxford Instruments) with a base temperature of $\sim\,\SI{270}{mK}$. The light-weight STM head and the cryostat cabling are designed similar to \cite{debuschewitzCompactVersatileScanning2007} to achieve high mechanical stability as well as high energy resolution.
\\
The tip, Fig.~\ref{fig:stm_setup}(1), is glued with silver conductive paint onto a copper slider - two copper half cylinders clamped to two sapphire rods -, and connected electrically to the voltage bias line via a $\SI{25}{\micro\meter}$ thin gold wire. The sapphire rods are glued to a single-tube four-quadrant piezo with a length of 1\," and a diameter of 0.25\," used for both: the approach in $z$-direction and the lateral scanning in the $x-y$ plane.
For the coarse approach we use the slip-stick mechanism \cite{rennerVerticalPiezoelectricInertial1990} by applying sawtooth voltage pulses with an amplitude of \SI{200}{\volt} to the $z$-electrode of the piezo tube. The possible lateral scan range by bending the piezo tube is estimated to be $\sim\,$\SI{2}{\micro\metre} in total at low temperature, as specified by the manufacturer.
\\
The sample is glued with silver conductive paint onto a copper sample plate (2),
that is screwed on top of two nanopositioners (3) and thermally anchored to the copper housing (5) with a copper braid. Using Kapton foil, the sample is insulated from ground.
To prevent adsorption and condensation of residual gas in the isolation vacuum onto the sample, it is heated during cooldown with a home-built heater (4). The latter consists of a manganin\textsuperscript{\textregistered} wire wound around a copper rod that is screwed onto the sample plate. For heating, a voltage of up to \SI{20}{\volt} is applied to the heater, elevating the sample temperature to a few K. 
Additionally to the thermometer at the $^3$He pot, we added a calibrated RuO$_x$ thermometer (6) closer to the sample.
\\
To improve thermal contact between the individual constituents of the STM head, all copper parts are gold plated and APIEZON\textsuperscript{\textregistered} N  grease is used at contact areas.
\\
The STM head is shielded against thermal radiation from the helium bath by copper screens that can be closed completely (7),  but also opened easily at room temperature for pre-adjusting the STM tip to the points of interest on the sample. The STM is controlled by a commercial RHK R9 controller with the I/V converter IVP-300 with $10^9$V/A . To increase bias voltage resolution we use a 1:100 resistive voltage divider at room temperature. 

\subsection{Filtering}
Apart from good thermal anchoring, it is necessary to filter the lines to isolate the signal from the room temperature (black-body) radiation and high-frequency noise that would affect the electronic temperature of the samples under investigation. Owing to the design of the insert, the first thermalization of the cables connecting room temperature with base temperature, is at the 1\,K stage (and not at the 4\,K flange as in many larger set-ups). Therefore, the thermal conductance and the total heat load has to be limited as much as possible, requiring a careful choice of the wiring material and filter concept.    

We use two complementary filter types: capillary filters that act as low-pass filters with attenuation above $\sim\,$\SI{900}{\mega\hertz}, and copper powder filters for high-frequency noise above $\sim\,$\SI{1}{\giga\hertz}.
\\
Copper powder filters consist of wires wound in a special geometry \cite{thalmannComparisonCryogenicLowpass2017} to compensate inductance, within a housing filled with copper powder with a grain size of around \SI{30}{\micro\metre}. The filtering concept was first introduced in 1987~\cite{martinisExperimentalTestsQuantum1987} and subsequently further developed~\cite{thalmannComparisonCryogenicLowpass2017}. It is based on the skin effect, which achieves strong attenuation by exploiting the large surface area of the oxidized copper powder. These filters also operate reliably at cryogenic temperatures.
\\
The current and bias lines are filtered twice with copper powder filters, first at 
the \SI{1}{\kelvin} stage and then at the \ce{^3He} stage, with each 
filter also serving as a thermal anchor at its respective stage. 
In \cite{thalmannComparisonCryogenicLowpass2017}, the amount of filters was reduced by using twisted pairs in the copper powder filters. However, such a scheme introduces capacitive coupling between the lines, thereby increasing noise. To minimize this crosstalk in the measurement lines, current and bias have their own dedicated filters at all stages.
\\
In addition, current and bias lines are filtered with capillary 
cables~\cite{thalmannComparisonCryogenicLowpass2017} from room 
temperature to the \SI{1}{\kelvin} stage. An insulated wire is inserted into a 
capillary filled with the conductive high-frequency absorber ECCOSHIELD\textsuperscript{\textregistered}. With that, the dielectric thickness is determined solely by the polyimide insulation of the wire yielding a high capacitance and thus strong attenuation at frequencies 
above $\sim\,$\SI{900}{\mega\hertz}.
\\
The lines for the piezos and nanopositioners are filtered with copper powder 
filters at the \ce{^3He} stage. For scan piezo, manganin wires are used instead 
of copper, as their low thermal conductivity reduces heat conduction from the hotter stages. Additional thermal anchoring for these lines is achieved above and 
below the \SI{1}{\kelvin} stage by winding the wires around copper rods tightly 
screwed to the cryostat.

\subsection{Nanopositioners}
The scan range of the piezo tube covers an area of $2\times2$\,$\upmu$m$^2$ at low temperature. Here, we added two non-magnetic and non-superconducting linear nanopositioners ANPx101 from attocube, made from CuBe, rotated by $\SI{90}{\degree}$ with respect to each other to move the sample in $x$- and $y$-direction, Fig.~\ref{fig:stm_setup} (3). Similar to our copper slider for the tip approach, they also operate based on a slip-stick mechanism and work at mK temperatures by adjusting the applied voltage. They have a lateral size of $\SI{24}{mm}\times \SI{24}{mm}$ and a travel range of $\SI{5}{mm}$ thus increasing the accessible area significantly.
\\
For the cabling of the nanopositioners, the cable resistance needs to be carefully chosen, as the positioners require high currents (0.5-1\,A at low temperatures). For this purpose, we use thin copper wires ($\SI{112}{\micro\metre}$, isolated), resulting in a total cable resistance of only $\approx \SI{7}{\ohm}$. For both positioners, one line is filtered with a copper powder filter at the $^3$He stage while the other one is directly connected to ground at low temperature.
\\
Since the nanopositioners' top plates are not connected to ground, the sample plate is also only poorly thermally coupled. To improve the sample's thermalization, the sample plate (2) is connected to the copper housing using a copper braid, see Fig.~\ref{fig:stm_setup}.\\
Moving the nanopositioners can increase the temperature by a few hundred mK, due to friction and Joule heating. Coarse positioning is thus better done while condensing the $^3$He with the system at 1.5\,K. The first alignment of the tip on top of the navigation structure is performed at room temperature, using a small optical microscope. After cooling the sample down to 1.5\,K, a second alignment is performed, moving to the area of interest so that after cooling to 300\,mK only minimal corrections are necessary. 

\section{Characterization of the system}
To characterize the STM equipped with nanopositioners and evaluate its energy resolution and electronic temperature, the differential conductance of a  system exhibiting low-energy excitations is measured. A very well understood system at low temperatures is the superconductor aluminium, the electronic properties of which are well described by the theory of Bardeen, Cooper and Schrieffer (BCS theory) \cite{bardeenTheorySuperconductivity1957} with the density of states given as
\begin{align}
    N_s(E) = \begin{cases}
        \frac{E}{\sqrt{E^2 - \Delta^2}} & \text{for } |E| > \Delta, \\
        0 & \text{for } |E| < \Delta.
    \end{cases}
\end{align}
\\
Using a PtIr tip, the density of states of the tip is constant and the differential conductance - measured with the lock-in technique - is then directly proportional to the density of states of the sample \cite{tinkhamIntroductionSuperconductivity2004, wiebe300MKUltrahigh2004}
\begin{align}
    \frac{dI}{dV} &\propto\int_{-\infty}^{\infty} N_s(E) \left[ - \frac{\partial f(E + eV)}{\partial (eV)}\right] \mathrm{d}E, \nonumber \\
    &\propto \int_{-\pi/2}^{\pi/2} \sin \alpha \cdot I(V + \sqrt{2}V_{mod} \sin \alpha) \mathrm{d}\alpha . \label{bcs_didv}
\end{align}

\begin{figure}[htbp]
	\centering
	\includegraphics[width=1.0\linewidth]{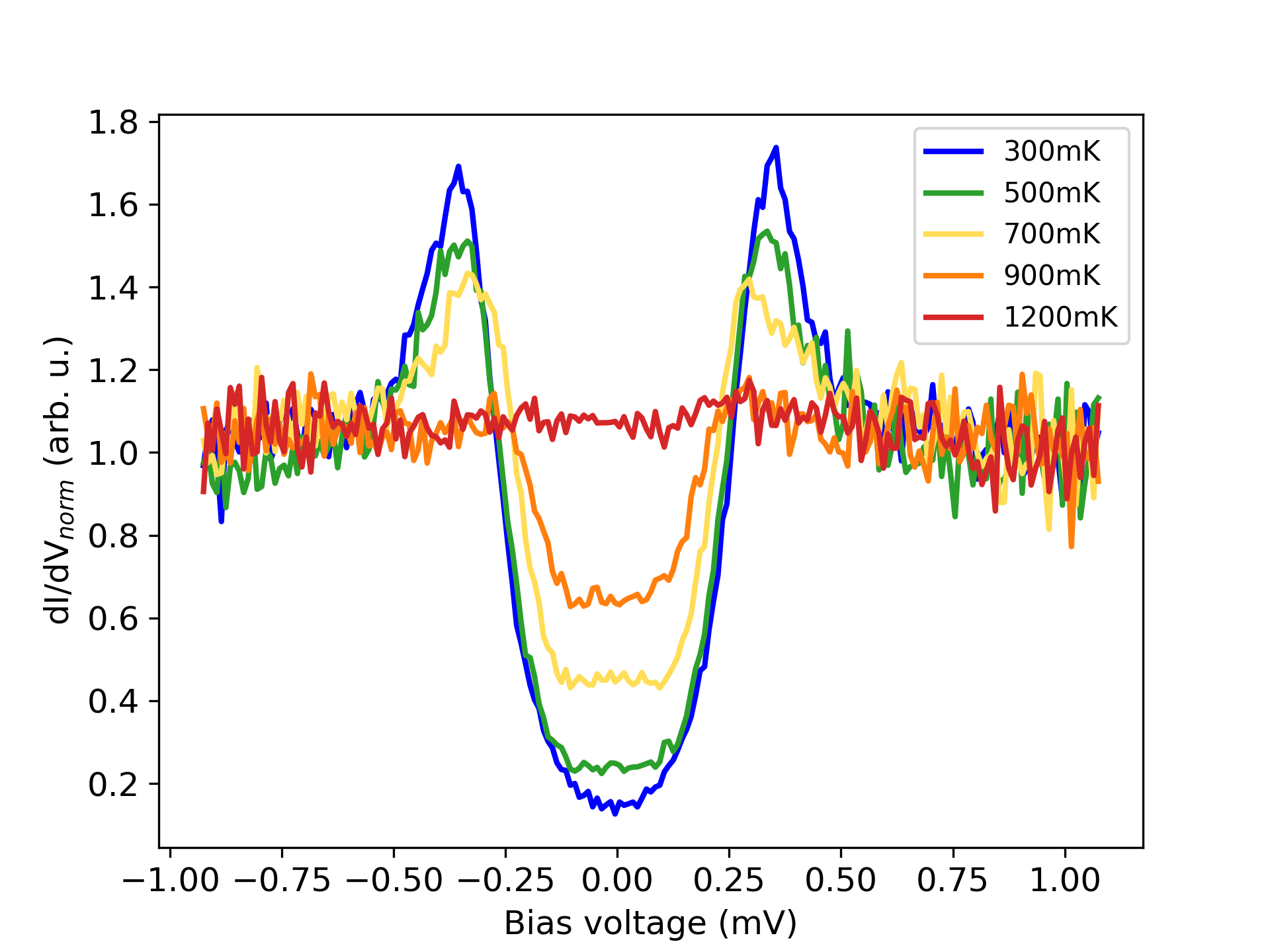}
	\caption{Measured differential conductance $dI/dV$ of an PtIr tip and an evaporated film with 200\,nm aluminium and 10\,nm copper for different temperatures. The spectra were recorded with a modulation voltage $V_{mod} = 30\,\upmu V$, a stabilization current $I = 400$\,pA at $V = 3$\,mV.}
	\label{fig:tempdependence}
\end{figure}

The measured differential conductance for different temperatures are shown in Fig.~\ref{fig:tempdependence} on a sample with 200\,nm Al and 10\,nm Cu on top to prevent oxidation and to assure clean tunneling conditions. The spectra were recorded with a modulation voltage $V_{mod} = 30\,\upmu V$ and a stabilization current $I = 400$\,pA at $V = 3$\,mV. The lines are experimental values at different temperatures, averaged over 10 measurements.

\section{NAVIGATION SYSTEM}
During cooldown, thermal drifts cause the tip position relative to the sample to shift in an uncontrolled manner. Since our setup provides no optical access, the exact tip position remains unknown at all times. To compensate for this drift, we integrated two nanopositioners that allow precise sample movement in the $x-y$ plane at cryogenic temperatures.
Navigating to points of interest requires the knowledge of the exact position of the tip and the required displacement. To this end, we developed a navigation system based on the sample topography. This approach allows reliable navigation to predefined regions of interest, such as nanoflakes or fabricated devices, where scanning tunneling spectroscopy measurements can be performed.

\subsection{Design}
\begin{figure}[htbp]
	\centering
	\includegraphics[width=0.8\linewidth]{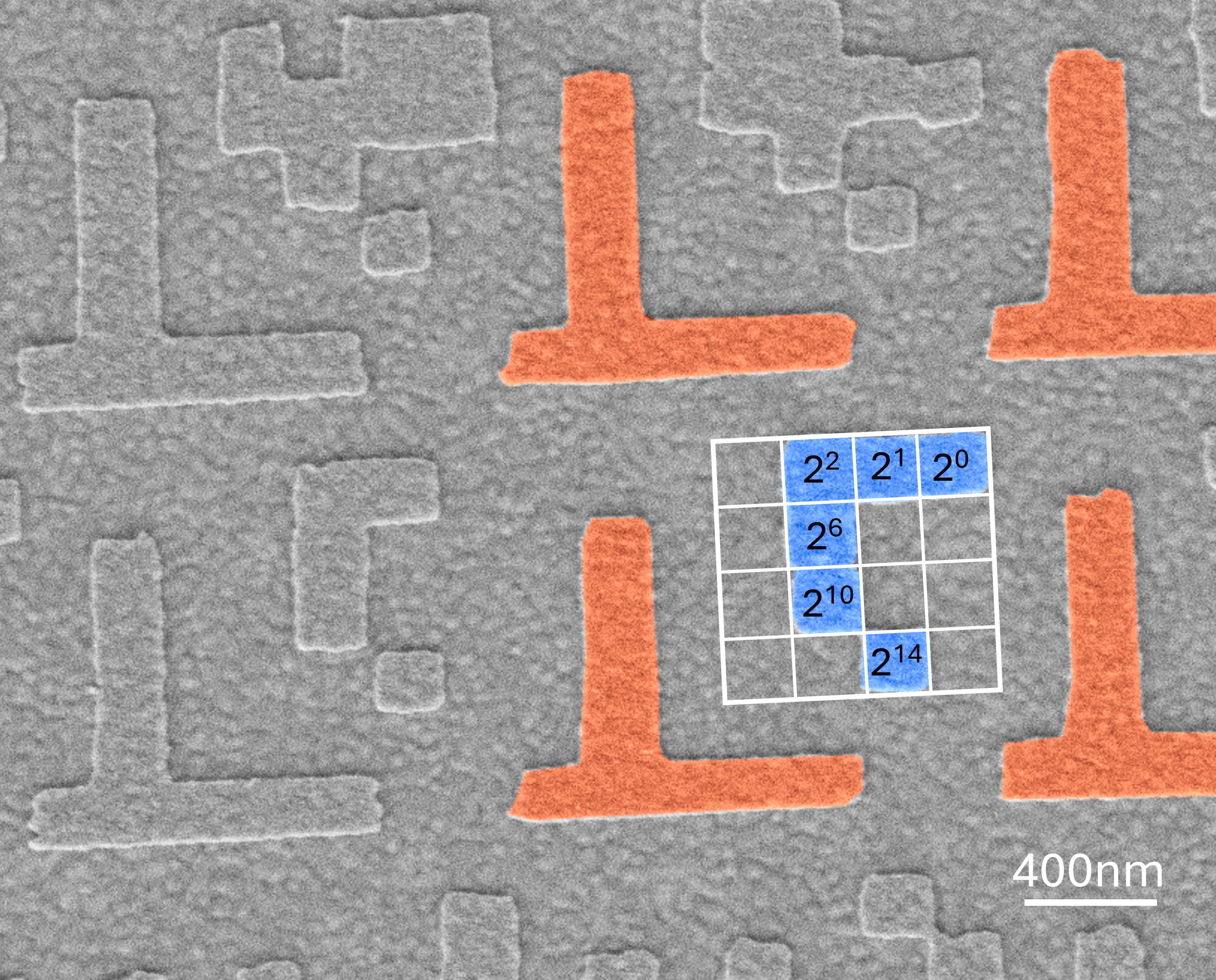}
	\caption{False-colored SEM image of a sample with the navigation system on top. The design is based on the dual system with a 4$\times$4 grid, shown in white, of 200\,nm$\times$200\,nm pixels that are either 1 (elevated by 10nm, blue) or 0 (no elevation). For a clear separation of grids, the border (orange) is also elevated and has an antisymmetric form in x and y for clear orientation of the pattern.}
	\label{fig:navidesign}
\end{figure}

An important design requirement is given by the limited scan range of the STM head. The design should be such that the relevant information - location of the tip and relative direction to the point of interest - can be extracted with one scan, thus not larger than 2$\upmu$m$\times$2$\upmu$m. Further, the absolute size of the navigation pattern should cover as much area as possible, to ensure the first-landing position of the tip after cooldown is within the patterned area. \\
For this purpose, we divide the sample into squares with an area of 1.2\,$\upmu$m$\times$1.2\,$\upmu$m each. In each square we write a number, denoting its position on the sample. To be able to cover a large area, we want to use as many numbers as possible and also as clear as possible to be readable with the STM. To this end, we use binary numbers and a grid of 4$\times$4 pixels, representing a dual number with 16 digits. Elevated pixels stand for a 1 while flat pixels are a 0.\\
In Fig.~\ref{fig:navidesign}, a false-colored SEM image of a sample with the navigation system on top is shown. The white grid shows the 4$\times$4 pixels with the 16 digits. The pixels marked in blue are elevated, thus forming the number 0100 0010 0010 0111. To the left, the previous number is shown, i.e., without the square encoding the $2^0$ being elevated. 
Each pixel has a size of 200\,nm$\times$200\,nm, giving a good resolution with our lithography system. The border structure, marked in orange, defines the start and end of the binary number and helps to distinguish them from each other. It is designed to be antisymmetric in $x$ and $y$ to distinguish different pattern orientations and also determine whether a scan is mirrored.
\\
With our setup, using 16 pixels with a size of 200\,nm$\times$200\,nm, an area of 350$\upmu$m$\times$350\,$\upmu$m can be covered on the sample. 
Depending on the scan area of the STM and the resolution of the fabrication setup, this area can be increased quite easily, e.g., by increasing the number of pixels. Also, this exact navigation pattern could be surrounded by a simpler design pointing in the direction of the pattern.

\subsection{Fabrication}
The navigation pattern is fabricated on \ce{SiO2}/Si substrates with 5\,nm Ti and \SI{20}{\nano \meter} Au evaporated on top to build a conductive sample. Using E-beam lithography and a polymethylmethacrylate as a resist, we can achieve resolutions of less than \SI{100}{\nano \meter}. Each pixel in the design has a size of 200\,nm x 200\,nm. Hence, one square with 16 pixels and the border has a size of $\SI{1.2}{\micro \meter} \times \SI{1.2}{\micro \meter}$ which still can be scanned by our STM head at low temperatures. After E-beam lithography, the substrates are developed in a solution of methyl isobutyl ketone and isopropanol. Then, there is another layer of Au evaporated on top building the navigation pattern. We tested different thicknesses of Au on top, between 5 and \SI{30}{\nano \meter}. \SI{15}{\nano \meter} is a good tradeoff between STM performance (if the steps are too high, there are many tip crashes) and visibility during optical alignment at room temperature. The PMMA layer is then removed with acetone so that the sample is ready for the STM.\\
An SEM image of a pattern prepared in this way is shown in Fig.~\ref{fig:navidesign}. 
This navigation pattern can now be universally exploited. With a transfer stage for 2D materials by HQ graphene, we can transfer nanoflakes on top of the pattern to measure in the STM but with another lithography step, one could also structure contact leads for transport measurements.

\subsection{Implementation}
\begin{figure}[htbp]
	\centering
	\includegraphics[width=1.0\linewidth]{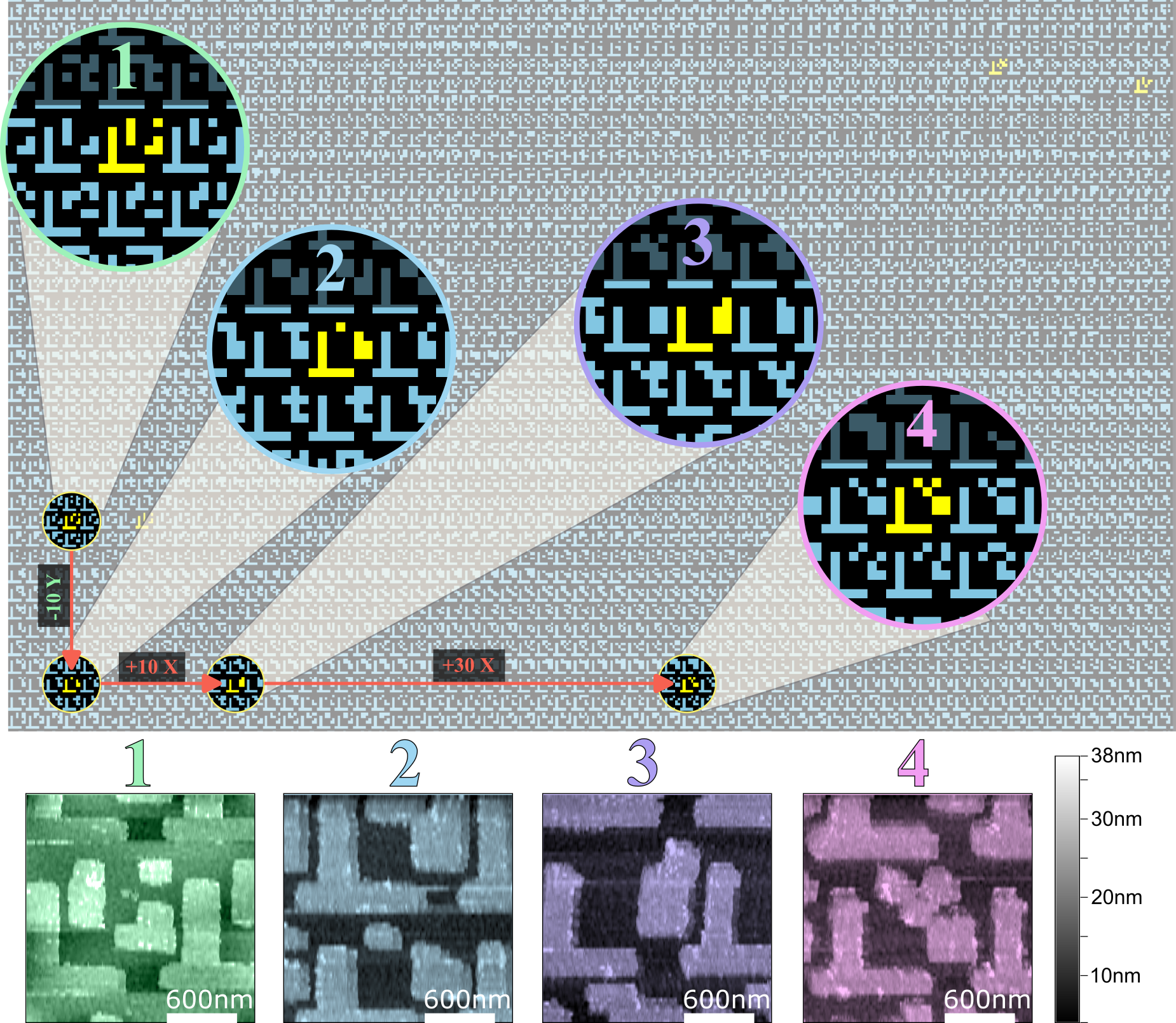}
	\caption{Topography-based navigation process. STM scans (1-4) at room temperature and the software (big image) showing the exact position in the pattern and therewith on the sample. Between the STM scans the attocube nanopositioners are moved.}
	\label{fig:navigationpathsmall}
\end{figure}

To read out the exact position in the pattern at low temperature, we scan an area of $\sim$2\,$\upmu$m and read out the 16 pixels of the 4x4 grid. Using a self-build program written with the GODOT\textsuperscript{\textregistered} Engine, we can deduce the exact position in the pattern and how to navigate to the point of interest. \\
Fig.~\ref{fig:navigationpathsmall} shows an example of a topography-based navigation process at room temperature. Scanning an area of 2\,$\upmu$m$\times$2\,$\upmu$m so that the border and at least one 4x4 grid is visible, one can read out the code and find the absolute position in the pattern using the self-built program (top of Fig.~\ref{fig:navigationpathsmall}). Then, after moving 10 steps with the $y$-nanopositioner another scan (2) is performed. To move to position (4) the $x$-nanopositioner is moved 40 steps.\\
At the region of interest, one would then cool down to 270\,mK to perform scanning tunneling spectroscopy.

\section{Conclusion}
Here we present a compact millikelvin scanning tunneling microscope (STM) operating at $\sim\,$270 mK that solves a key practical challenge: reliably locating and navigating to micrometer-sized regions of interest on a sample without optical access at low temperatures.
\\
The setup builds on an existing compact lightweight STM design and adds two piezoelectric nanopositioners that extend the accessible sample area well beyond the $~ \SI{2}{\micro\meter}$ scan range of the piezo tube alone. To achieve good electronic temperature and energy resolution, we use an advanced filtering concept at multiple cryostat stages to reduce noise and secure a good thermalization of the electronic systems.  The instrument's performance is verified by measuring the superconducting gap of aluminum and comparing it to the BCS theory.
\\
The central innovation is a topography-based navigation system. A binary-encoded pattern of gold pixels (4$\times$4 grids of 200\,nm $\times$ 200\,nm pixels) is lithographically fabricated on the substrate, encoding absolute position coordinates across a 350\,$\upmu$m $\times$ 350\,$\upmu$m area. By scanning a 2$\upmu$m$\times$2$\upmu$m region and reading the binary grid, custom software identifies the tip's exact location on the sample. The nanopositioners can then move the sample to any target position to perform STS with a high energy resolution of $\SI{20}{\micro\electronvolt}$. The design requires only a single lithography step and is compatible with a wide range of materials and STM systems.
\\
It is possible to place multiple devices or nanostructures on a single sample and reliably return to any position of interest. The primary limitation is that the tip must be brought into tunneling range to read the coordinate pattern, risking the tip to crash during scanning. This risk is reduced by the small step size of 15\,nm in the pattern and depends also on the scan speed. This is in contrast to capacitance-based approaches, where the navigation is performed far away from the sample. Future work could address this by combining the here presented topographic approach with a coarse capacitive detection step.

\begin{acknowledgments}
We thank A. Di Bernardo for helpful discussions and A. Fischer for technical support. Further, we acknowledge the nano.lab of the University of Konstanz and thank its staff, M. Hagner, S. Haus and A. Zuschlag, for their expert advice.  We also gratefully acknowledge financial support from the Deutsche Forschungsgemeinschaft (DFG; German Research Foundation) via project No. 493158779 as part of the collaborative research project SFB F 86 Q-M\&S funded by the Austrian ScienceFund (FWF, project number LAP 8610-N).
\end{acknowledgments}



\bibliography{Zotero26_paper2}

\end{document}